\documentclass[num-refs]{nbdt-article}

\usepackage{siunitx}

\pdfoutput=1

\papertype{Original Article}

\title{On the Subspace Invariance of Population Responses}

\author[1]{Elaine Tring}
\author[1,2]{Dario L. Ringach}

\affil[1]{Department of Neurobiology, David Geffen School of Medicine, University of California, Los Angeles, CA 90095, USA.}
\affil[2]{Department of Psychology, University of California, Los Angeles, CA 90095, USA.}

\corraddress{Dario L. Ringach. Department of Neurobiology, David Geffen School of Medicine, University of California, Los Angeles, CA 90095, USA.}
\corremail{dario@ucla.edu}

\fundinginfo{This study was funded by grants NIH EY018322 and EB022915 to DLR.}

\runningauthor{Tring {\em at al}}

\begin{document}

\maketitle

\begin{abstract}
In cat visual cortex, the response of a neural population to the linear combination of two sinusoidal gratings (a plaid) can be well approximated by a weighted sum of the population responses to the individual gratings --- a property we refer to as {\em subspace invariance}. We tested subspace invariance in mouse primary visual cortex by measuring the angle between the population response to a plaid and the plane spanned by the population responses to its individual components. We found robust violations of subspace invariance arising from a strong, negative correlation between the responses of neurons to individual gratings and their responses to the plaid. Contrast invariance, a special case of subspace invariance, also failed. The responses of some neurons decreased with increasing contrast, while others increased. Altogether the data show that subspace and contrast invariance do not hold in mouse primary visual cortex. These findings rule out some models of population coding, including vector averaging, some versions of normalization and temporal multiplexing.

\end{abstract}

\section{Introduction}
How do population of neurons in primary visual cortex encode the visual input \cite{RN2213,RN526,RN1038,RN672,RN82,RN793,RN2214}? Can we predict the response of a neural population to complex visual stimuli from the measurement of its responses to simple ones? These are fundamental questions in systems neuroscience. Previous work measured the response of neural populations to gratings of different contrasts and the plaids that result from their superposition \cite{RN820,RN2400}. It was reported that the mean population response to plaids could be described by a linear mixture of the mean responses to the individual component gratings. Moreover, the mixing coefficients were explained by a normalization model \cite{RN2395}. The model provided a satisfactory account for the averaging behavior observed when gratings had similar contrasts and the gradual transition to a winner-take-all regime as the contrasts became increasingly different \cite{RN820}. The normalization model has also been shown to explain cross-orientation suppression \cite{RN726,RN848}.

\begin{figure}[h]
\centering
\includegraphics[width=7cm]{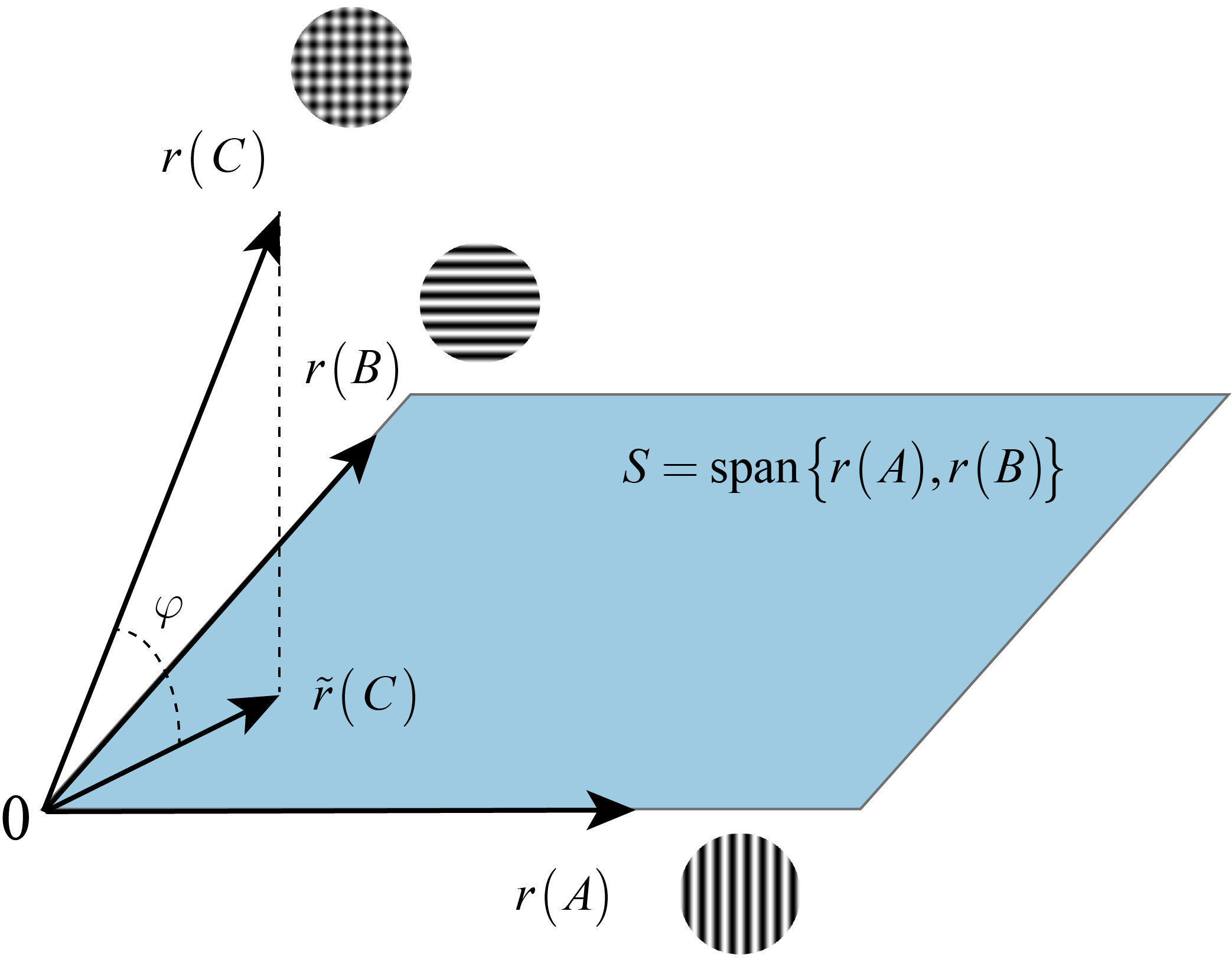}
\justify
\caption{Subspace invariance of population responses. The vectors $r(A)$, $r(B)$, and $r(C)$ represent the mean responses of a population to two individual gratings and the plaid resulting from their combination respectively. A population response is said to satisfy subspace invariance if the response to a linear combination of stimuli lies in the space spanned by the population responses to the individual stimuli. Departures from subspace invariance can be measured by the deviation angle $\varphi$ between the actual population response to a linear combination and the best approximation resulting by a linear combination of the responses to each of the components $\tilde{r}(C)$.  A system is subspace invariant if $\varphi \approx 0$.}
\end{figure}

The normalization model applied to neural population responses is an example of a system that satisfies {\em subspace invariance}.  In other words, given $C = \alpha A + \beta B$, subspace invariance is satisfied if $r(C) \in \textrm{span}\{r(A),r(B)\}$. One way to test for subspace invariance is to measure the angle $\varphi$ between $r(C)$ and its best approximation lying in the plane $S = \textrm{span}\{r(A),r(B)\}$ (\textbf{Fig 1}), which is predicted to be near zero. Of course, the optimal approximation of $r(C)$ in the mean-squared sense is the projection of $r(C)$ onto $S$, which we denote by $\tilde{r}(C) = P_S(r(C))$. It is easy to show that if a system satisfies subspace invariance, then it  also satisfies {\em contrast invariance} \cite{RN2244,RN587,RN843,RN830,RN580}.In other words, $r(\alpha A) \in \textrm{span}\{r(A)\} = C_\alpha r(A)$ for some constant of proportionality $C_\alpha$. In other words, scaling the contrast of a stimulus only changes the amplitude of the population vector leaving its direction unchanged. 

It is important to emphasize at the outset that evaluating contrast invariance at the population level is not the same as doing so for single cells \cite{RN820}. A single neuron is contrast invariant if its response can be factorized as $r(c,\theta) = f(c) g(\theta)$, where $c$ is the contrast of the stimulus with a spatial structure described by the parameter $\theta$.  The function $f(c)$ is called the {\em contrast response function} of the neuron, and $g(\theta)$ is called its {\em tuning function}. Assume we have a population of cells each of which is contrast invariant.  Each cell has its own contrast response and tuning functions. The population response is a vector $r(c,\theta) = (f_1(c)g_1(\theta), \ldots ,f_N(c) g_N(\theta))$. We say the {\em population} is contrast invariant if the direction of the response vector does not change with contrast, meaning that  $r(c,\theta) = f(c) (g_1(\theta), \ldots , g_N(\theta))$. Clearly, this can only occur if the contrast response functions for all neurons are the same $f(c) \equiv f_1(c) = f_2(c) = \ldots = f_N(c).$ Thus, a set of neurons, each of which is contrast invariant, is not necessarily contrast invariant at the population level. In fact, we know there is substantial variability in the shape of contrast response functions in cortical neurons \cite{RN2268,RN2269,RN2435}. Thus, in general, one may expect contrast invariance at the population level to fail. 

\section{Methods}

\subsection{Animals} 
All procedures were approved by UCLA’s Office of Animal Research Oversight (the Institutional Animal Care and Use Committee) and were in accord with guidelines set by the US National Institutes of Health. A total of 12 mice, both male (3) and female (9), aged P35-56, were used. All these animals were from a TRE-GCaMP6s line G6s2 (Jackson Lab), where GCaMP6s is regulated by the tetracycline-responsive regulatory element (tetO). Mice were housed in groups of 2-3, in reversed light cycle. Animals were naïve subjects with no prior history of participation in research studies. We imaged 51 different populations to obtain the data discussed in this study.

\subsection{Surgery} 
Carprofen and buprenorphine analgesia were administered pre-operatively. Mice were then anesthetized with isoflurane (4-5\% induction; 1.5-2\% surgery). Core body temperature was maintained at 37.5C using a feedback heating system. Eyes were coated with a thin layer of ophthalmic ointment to prevent desiccation. Anesthetized mice were mounted in a stereotaxic apparatus. Blunt ear bars were placed in the external auditory meatus to immobilize the head. A portion of the scalp overlying the two hemispheres of the cortex (approximately 8mm by 6mm) was then removed to expose the underlying skull. After the skull is exposed, it was dried and covered by a thin layer of Vetbond. After the Vetbond dries (15 min) it provides a stable and solid surface to affix an aluminum bracket (a head holder) with dental acrylic. The bracket is then affixed to the skull and the margins sealed with Vetbond and dental acrylic to prevent infections. Details on the implantation of cranial windows can be found elsewhere \cite{RN884}.

\subsection{Imaging}

We began imaging sessions 4-5 days after surgery. We used a resonant, two-photon microscope (Neurolabware, Los Angeles, CA) controlled by Scanbox acquisition software and electronics (Scanbox, Los Angeles, CA). The light source was a Coherent Chameleon Ultra II laser (Coherent Inc, Santa Clara, CA) running at 920nm. The objective was an x16 water immersion lens (Nikon, 0.8NA, 3mm working distance). The microscope frame rate was 15.6Hz (512 lines with a resonant mirror at 8kHz). Images were captured at an average depth of $220 \mu \textrm{m}$. 

\subsection{Visual stimulation}

In all experiments we used a BenQ XL2720Z screen which measured 60 cm by 34 cm and was viewed at 20 cm distance, subtending 112 x 80 degrees of visual angle. The screen was calibrated using a Spectrascan PR-655 spectro-radiometer (Jadak, Syracuse, NY), and the result used to generate the appropriate gamma corrections for the red, green and blue components via an nVidia Quadro K4000 graphics card. The contrast of individual sinusoidal gratings was 70\%. The plaid was the sum of two gratings with a contrast of $70\% / \sqrt{2} = 50\%$, to ensure the same effective contrast as the individual gratings. The spatial frequencies of the components were always the same to each other, and ranged from 0.03 to 0.06 cpd. In a series of experiments conducted in 3 animals we also measured the full orientation and spatial frequency tuning of neurons by presenting a sequence of gratings with randomized orientations and spatial frequencies as described in detail elsewhere \cite{RN2373,RN2374,RN2369}. Visual stimuli were generated in real-time by a Processing sketch using OpenGL shaders (see \url{http://processing.org}) Transistor-transistor logic (TTL) signals generated by the stimulus computer were sampled by the microscope and time-stamped with the frame and line number that being scanned at that time. The time-stamps provided the synchronization between visual stimulation and imaging data. The approximate locations of the receptive fields of the population were estimated by an automated process where localized (5x5 deg), flickering checkerboards patches, appeared at randomized locations within the screen. This experiment was run at the beginning of each imaging session to align the centers of the receptive fields on the center of the monitor. We imaged the monocular region of V1 in the left hemisphere. The receptive fields of neurons were centered around 20 to 35 deg in azimuth and 0 to 20 deg in elevation on the right visual hemifield. 

\begin{figure}[ht]
\centering
\includegraphics[width=14cm]{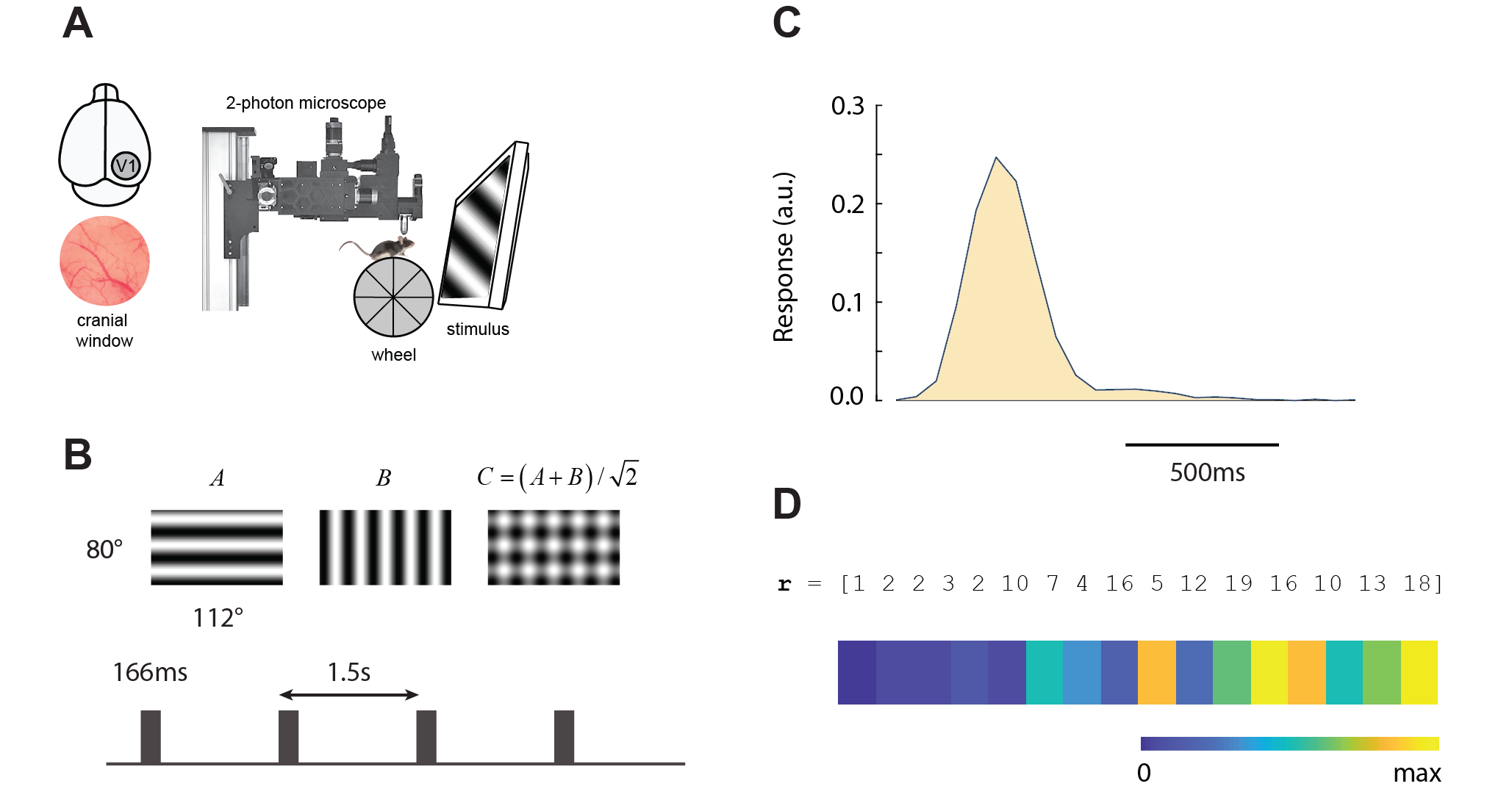}
\justify
\caption{Experimental setup. (\textbf{A}) A cranial window was implanted on top of primary visual cortex (area V1). Awake mice were imaged during the presentation of visual stimuli. (\textbf{B}) Visual stimuli consisted of two orthogonal sinusoidal gratings and a plaid resulting from their combination. Patterns were flashed briefly in random order. (\textbf{C}) Typical, average temporal response of a single cell to one of the stimuli. The response returns to baseline by the end of the 1.5 trial. The response of the cell is summarized by its integrated activity over time (shaded area). (\textbf{D}) The population response vector comprised of the activity of all cells in the population is represented as “barcode”, where response rate is represented by a colormap. Blue hues represent low firing rates and yellow hues represent high firing rates.  Barcodes are normalized between zero and the maximal response of the population.}
\end{figure}

\subsection{Image Processing}

Image processing: The image processing pipeline was the same as described in detail elsewhere \cite{RN2373}. Briefly, calcium images were aligned to correct for motion artifacts. We then used a Matlab graphical user interface (GUI) tool developed in our laboratory to define regions of interest corresponding to putative cell bodies manually. Following segmentation, we extracted signals by computing the mean of the calcium fluorescence within each region of interest and discounting the signals from the nearby neuropil. Spikes were then estimated via deconvolution \cite{RN2172}. All the responses analyzed here are based on the estimated spike rates.

\subsection{Data Selection}

The mean temporal response of each cell for each condition was computed (\textbf{Fig 2C}). We considered a cell to have generated a significant response to a given condition (one of the two gratings or the plaid) if the peak response relative to the baseline was larger than eight. We verified our findings are robust to changes in this criterion. We defined cell populations as the set of all cells which generated a significant response for at least one condition, all other neurons were discarded. In some experiments, we also measured the joint tuning in the orientation and spatial-frequency domain of the same population of neurons. Kernels with a signal-to-noise ratio (SNR) above 2.5 were selected for further analysis. The SNR was defined by the ratio of the standard deviation of the kernel at the optimal delay time to the standard deviation of the kernel at times less than zero.  

\subsection{Data Availability} The data analyzed in this study are available in Figshare \url{https://figshare.com/articles/_/7094570}.

\section{Results}

We measured the activity of neural populations in primary visual cortex of awake, adult mice of both genders, expressing a calcium indicator (GCaMP6s) during exposure to computer-controlled visual stimulation (\textbf{Fig 2A}). Mice were head restrained, but otherwise free to walk, rest or groom on a 3D printed wheel. The stimulus sequence contained three different patterns: two sinusoidal gratings with the same spatial frequency but orthogonal orientations (patterns $A$ and $B$), and a plaid resulting from their sum scaled to keep the same effective contrast (pattern $C=(A+B)/\sqrt{2}$) (\textbf{Fig 2B}). In each trial, one of the patterns was selected at random with equal probability and flashed for 166ms, with a stimulus-onset asynchrony of 1.5s (\textbf{Fig 2B}). The responses of individual cells to such brief stimuli return to baseline by the end of the trial (\textbf{Fig 2C}). We define the response of a cell as its integrated activity over the entire trial (\textbf{Fig 2C}, shaded area). We ran 30 min long sessions for a total of 1200 trials (400 average presentations per pattern). We computed the average response for each visual pattern across trials, $r(A)$, $r(B)$, and $r(C)$. The dimension of the population responses equals the number of cells in the population after selecting the ones that responded significantly to at least one stimulus pattern (see \textbf{Methods}). 

It is convenient to represent a population response as a “barcode” where the response of each cell is mapped to a normalized colormap (\textbf{Fig 2C}). To make it easier to visually interpret population barcodes, we ordered the cells according to their relative preference for the two grating stimuli. Denote by $r_i(A)$ and $r_i(B)$ the mean response of the $i-th$ cell in the population to patterns $A$ and $B$ respectively. Then, barcodes display the population response when cells are ordered from left to right in ascending order of $r_i (A)-r_i (B)$. In this representation, cells with preference for $A$ over $B$ will be on the right hand side of the barcode, while cells with a preference for $B$ over $A$ will be on the left. Cells near the middle respond equally well to both stimuli --- note that those responses may be equally high or equally low. 

\begin{figure}[ht]
\centering
\includegraphics[width=12cm]{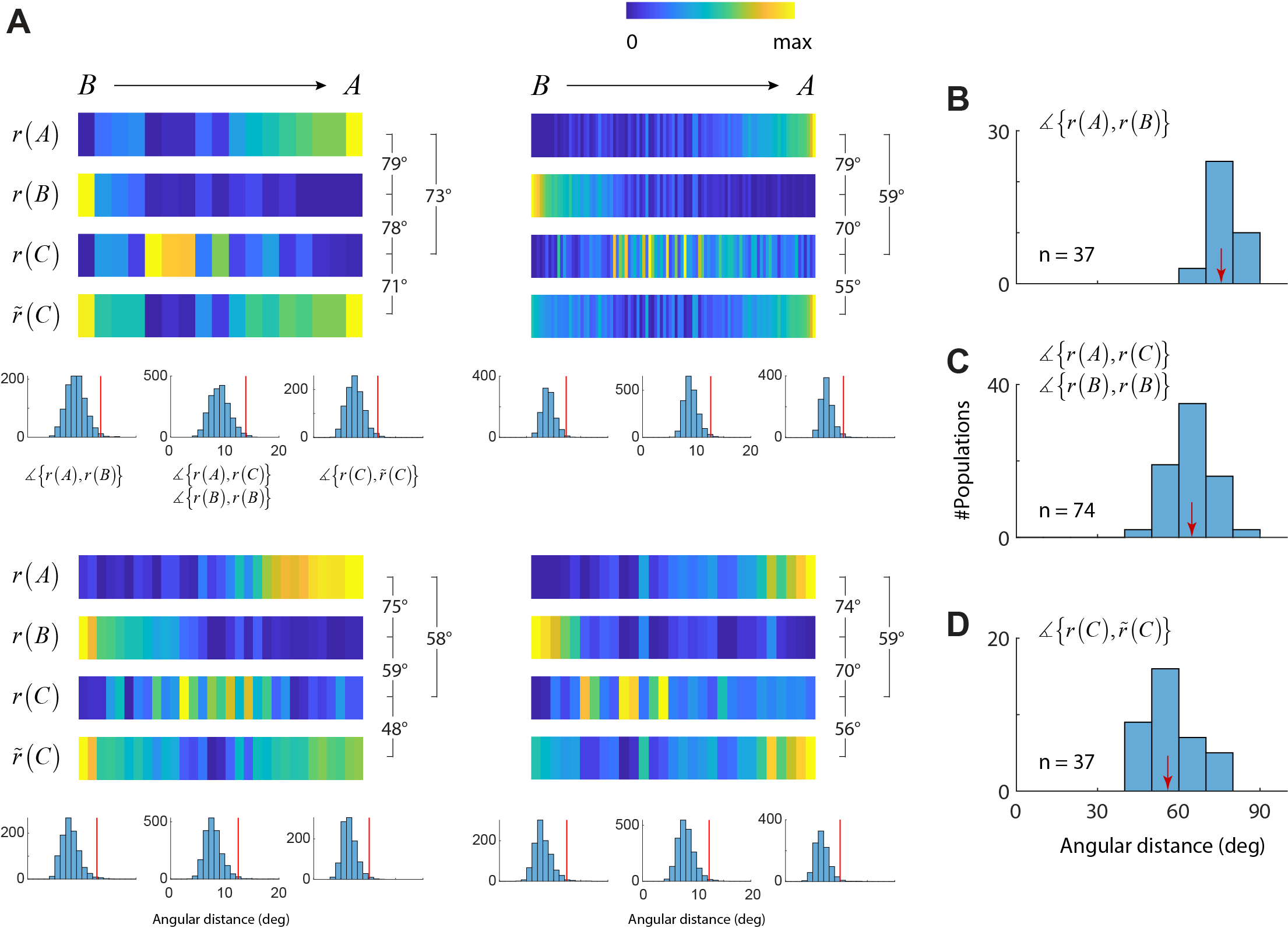}
\justify
\caption{Testing subspace invariance in mouse V1. (\textbf{A}) Each panel shows the results from separate experiments. Within each panel, stacked barcodes display the mean population responses to the gratings, $r(A)$ and $r(B)$, the mean response to the plaid, $r(C)$, and the best approximation to the population response to the plaid that lies in the plane spanned by $r(A)$ and $r(B)$, denoted by $\tilde{r}(C)$. The number between the brackets indicate the angular distance between the different population vectors. Histograms below the panels show the expected distribution of angular distances under random permutation of the classes (n=1000 simulations), with the vertical red line indicating the 99\% of the distribution. Barcodes in this and other figures are normalized between 0 and their maximal responses separately. (\textbf{B}) Distribution of angular distance between mean population responses to the sinusoidal gratings across experiments. (\textbf{C}) Distribution of angular distance between mean population responses to the sinusoidal gratings and the plaid. (\textbf{D}) Distribution of angular distance between the response to the plaid and the best approximation in the plane spanned by the grating responses. (\textbf{B-D}) Red arrows indicate median angular distance in each case.
}
\end{figure}

For each experimental session, we visualized the results by stacking the barcodes representing the mean population responses to each component grating $r(A)$ and $r(B)$, the mean population response to the plaid, $r(C)$, and the best approximation to $r(C)$ attainable by a linear combination of $r(A)$ and $r(B)$, which we denoted by $\tilde{r}(C)$ (\textbf{Fig 3A}). Due to our choice of cell ordering, the barcode for $r(A)$ shows the most active cells are on the right, while the most active cells in the barcode for $r(B)$ are on the left. Cells that are active during the presentation of a grating appear not to be respond strongly to the plaid, as both ends of the barcode for the plaid are dominated by blue hues, representing to low firing rates. In contrast, cells that respond strongly to the plaid are positioned somewhere in the middle of the barcode for $r(C)$. These cells do not seem to respond strongly to the individual components. Simple visual inspection reveals that the population response to a plaid is very different from its best approximation obtained as a linear combination of the population responses to the components, violating the prediction of subspace invariance (\textbf{Fig 3A}). In any one experiment the mutual distances between the population vectors was compared to those obtained by randomly permuting the labels of the stimulus classes (\textbf{Fig 3A}, histograms). We ran 1000 random permutations and computed the angular distance at the 0.01 level in each case, which was typically around 15 deg or less (\textbf{Fig 3A}, red vertical line in histograms).  In contrast, the mutual distances between the actual response were all larger than 40 deg. In other words, all angular distances computed in single experiments were {\em all} significantly larger than expected by chance at the $p=0.01$ level.

Across all experiments, the mean angular distance between the population responses to the gratings was large, with a median of 75.4 deg (\textbf{Fig 3B}). The median angular distance between the population response to the gratings and the plaid stimulus was 64.2 deg (\textbf{Fig 3C}). Finally, failure of subspace invariance is implied by a median distance between the population response to the plaid and its projection on the subspace spanned by the responses to gratings, of 56.5 deg. The medians of the distributions were all statistically different from each other at the $p<10^{-4}$ level (rank-sum test).

\begin{figure}[h]
\centering
\includegraphics[width=10cm]{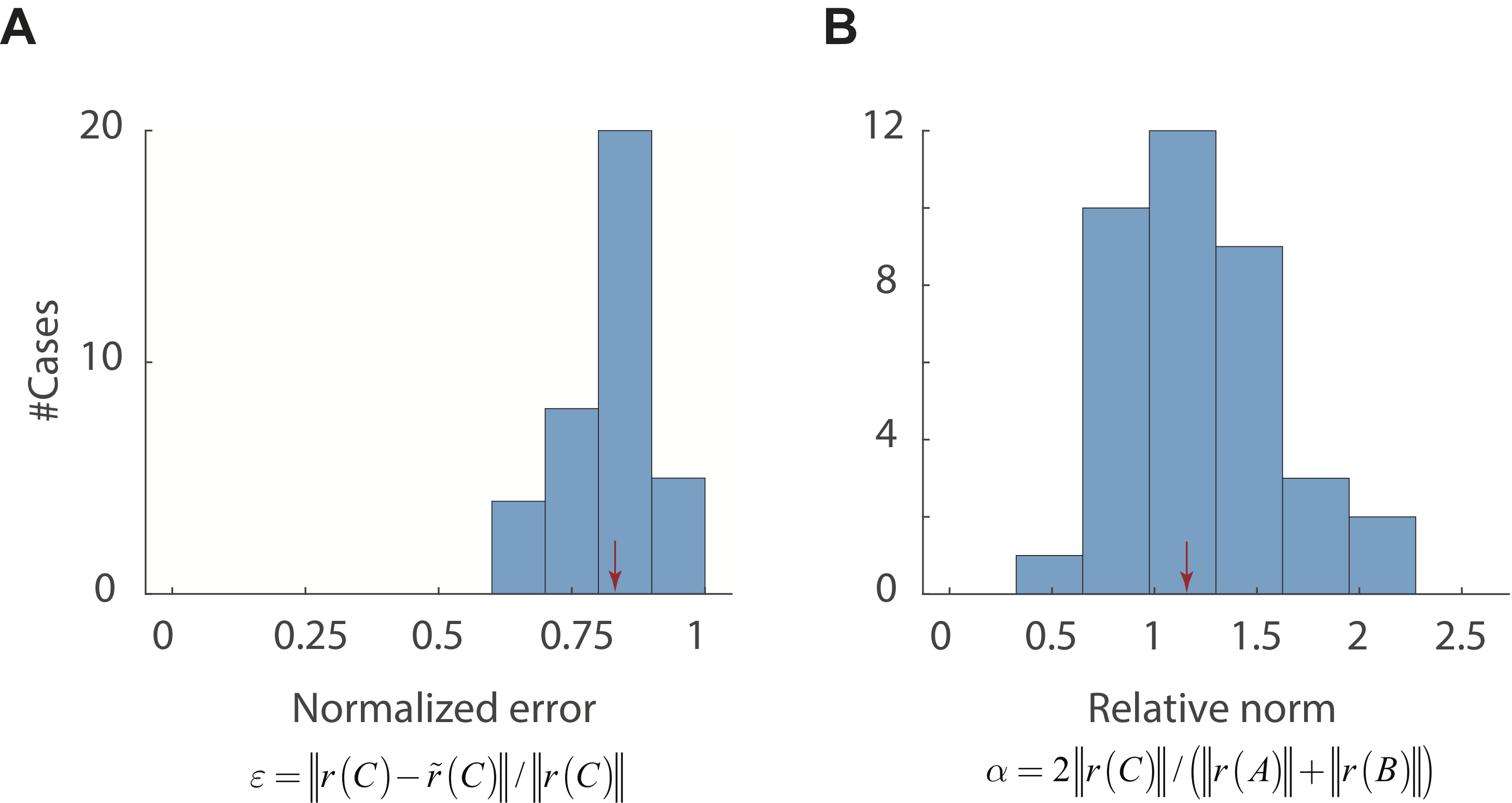}
\justify
\caption{Normalized error and relative magntitudes. (\textbf{A}) Histogram of normalized error for all experiments in the dataset. (\textbf{B}) Relative magnitude of the plaid population responses with respect to those evoked by gratings. 
}
\end{figure}

An alternative way to express the deviations from subspace invariance is to calculate the relative error, defined as the norm of the difference between the actual plaid response and its projection, normalized by the norm of the plaid response, $\epsilon = \| r(C) - \tilde{r}(C) \| / \| r(C) \|$ (\textbf{Fig 4A}). The normalized error is relatively large, with a median of 0.83. These data are the same as shown in \textbf{Fig 3D}, as one can easily see that $\epsilon = \sin \varphi $ (\textbf{Fig 1}). These large deviations are not the result of weak population responses to the plaid, as the magnitude of the population responses to the gratings and the plaids were comparable ({\textbf{Fig 4B}). In fact, plaids generated responses that were 1.15 times larger than the mean norm of the responses to the gratings. 

To summarized our data, we computed an “average barcode” across all experiments. First, we normalized the width of the barcodes to one and considered the values of the barcode as samples of a function in the $[0,1]$ interval. We then interpolated these data at 32 equally spaced points and averaged the results for each stimulus class to obtain average barcodes and angular distances (\textbf{Fig 5A}). These composite barcodes highlight the difference in the population responses for each stimulus condition. It also accentuates one key feature of the data already noted in the individual cases (\textbf{Fig 3A}) --- namely, cells in the population that respond to the plaids do not respond to the individual components, while cells that respond to the individual components do not respond to the plaid. This is the fundamental reason why it is not possible to approximate the population response to the plaid as a linear combination of the responses to the gratings. This effect, which is so salient in the average data, is also present if we look at the responses of individual cells (\textbf{Fig 5B}). The scatterplot shows the average response of a cell to the plaid, $r_i (C)$, against its mean response to the gratings,  $(r_i (A)+r_i (B))/2$. All individual cells recorded are included in this analysis. The responses of individual cells to plaids and gratings are negatively correlated (n=1035, r=-0.62, $p< 10^{-100}$).

\begin{figure}[h]
\centering
\includegraphics[width=12cm]{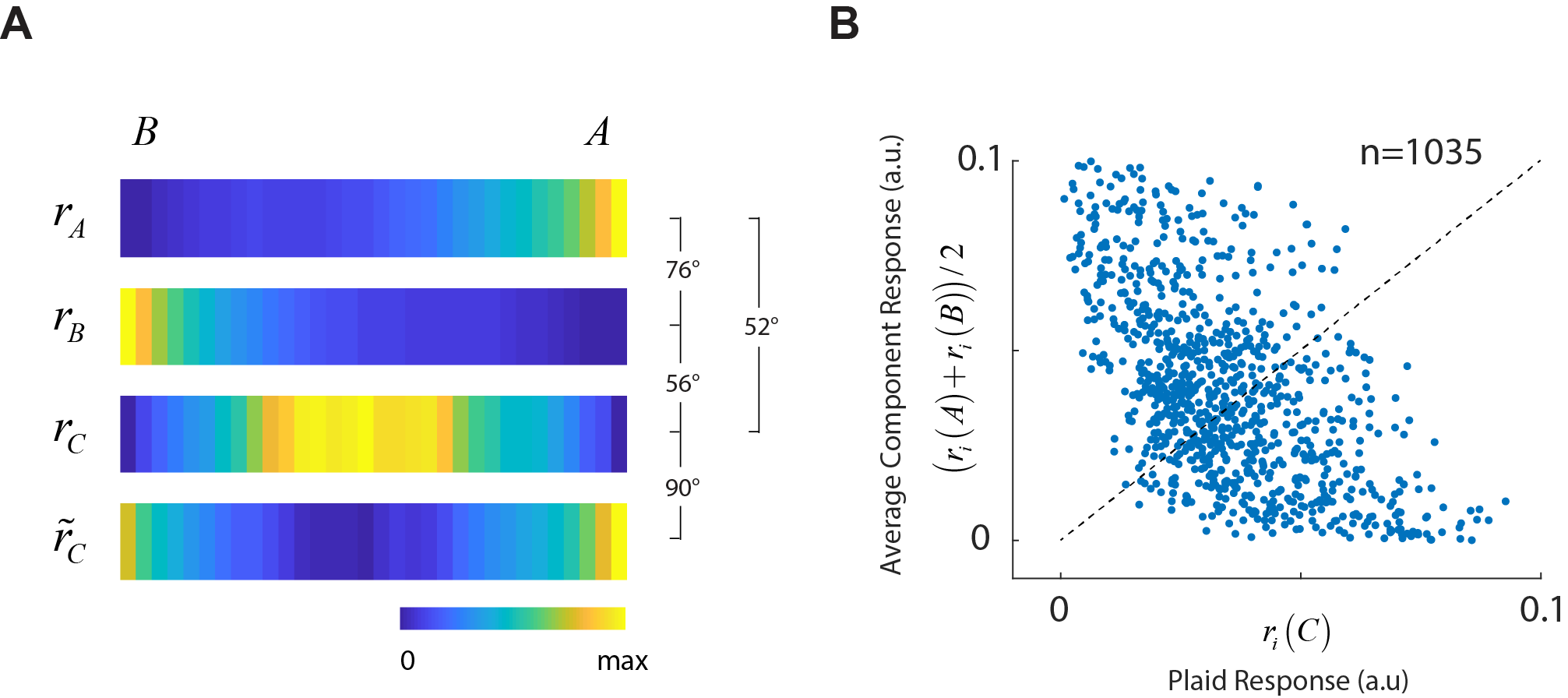}
\justify
\caption{Average barcodes and behavior of single neurons. (\textbf{A}) Average barcodes across all our experiments. As before cells are ordered according to their preference between gratings A and B. The average barcodes highlight the differences between the population responses to gratings and plaids. The best approximation to the plaid population from linear combination to the average response to gratings is poor. (\textbf{B}) A scatterplot of plaid versus average responses to gratings for each individual cell shows a statistically significant, and strong anticorrelation. Dotted line represents the unity line. 
}
\end{figure}

\begin{figure}[h]
\centering
\includegraphics[width=6cm]{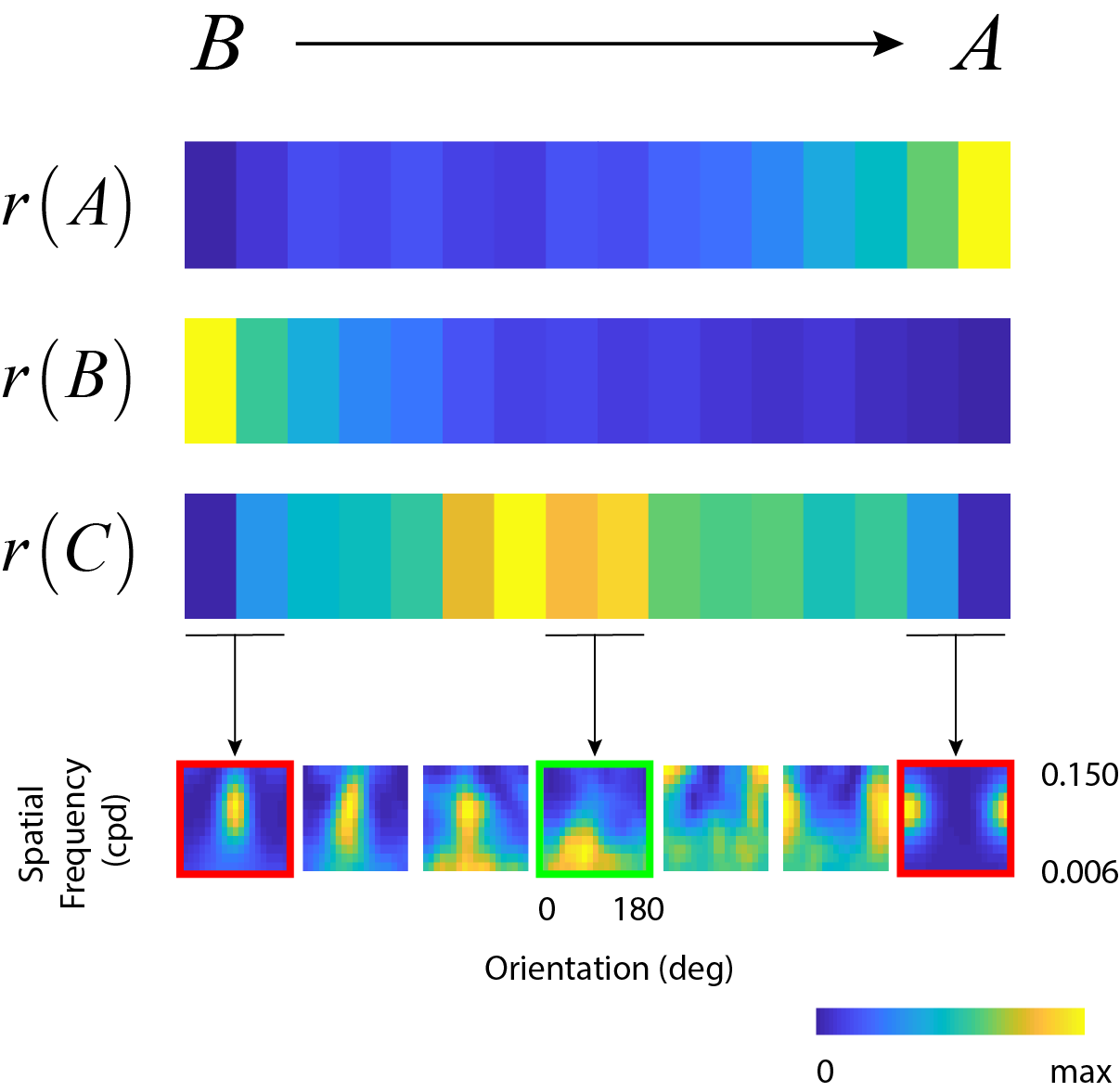}
\justify
\caption{Relationship between tuning in the Fourier domain and component/plaid responses. Cells which respond to each component, but not the plaid, show sharp tuning for orientation and spatial frequency, as shown by the leftmost and rightmost tuning kernels. The average tuning for cells which respond best to the plaids, in contrast, shows a more broadly tuned response for orientation and a preference for lower spatial frequencies, as depicted by the average tuning kernel for cells in the middle of the barcode. Note that spatial frequency is a logarithmic axis in these tuning kernels.
}
\end{figure}

What makes some cells respond to plaids and not gratings?  To explore this question we repeated the same experiments but, this time, we also obtained the joint tuning for orientation and spatial frequency of cells in the population.  These data were obtained in a total of 15 optical planes from 3 mice. In these experiments, we kept the components fixed at 0 and 90 deg (vertical and horizontal) and used a fixed spatial frequency of 0.04 cpd. 

As expected, we find that the average tuning for cells that preferentially respond to one component and not other are well tuned in orientation and spatial frequency, with their orientation preference matching that of the components (\textbf{Fig 6}, kernels with red outline).  Interestingly, cells that respond better to the plaids than the gratings, which are located near the middle of the barcode, display an average tuning that is more broadly tuned for orientation and low-pass in spatial frequency (\textbf{Fig 6}, kernel with green outline). These results imply that stimulation with a simultaneous pair of grating components do not merely engage the population of cells they activate individually, as would be predicted by subspace invariance. Instead, the most active cells are those that tend to have broad tuning for orientation and tuned to low spatial frequencies.

Finally, we tested contrast invariance at the population level by measuring the population response to a single grating with a fixed orientation and spatial frequency and two different levels of contrast, low (15\%) and high (80\%) in 4 optical planes from 2 mice. In this case, we created population barcodes by ordering cells according to the difference in their response between high and low contrast. The angular distance between responses to low and high contrast grating were statistically different (\textbf{Fig 7A}).
The average barcode averaged across experiments shows that some cells respond better to a low contrast stimulus than a high contrast one ({\textbf{Fig 7BC}).  In other words, such cells see their activity suppressed as the contrast of the stimulus increases. The existence of these cells may not be entirely surprising, as they are already present in the LGN \cite{RN2151}. These results indicate a failure of contrast invariance at the population level.

\begin{figure}[h]
\centering
\includegraphics[width=13cm]{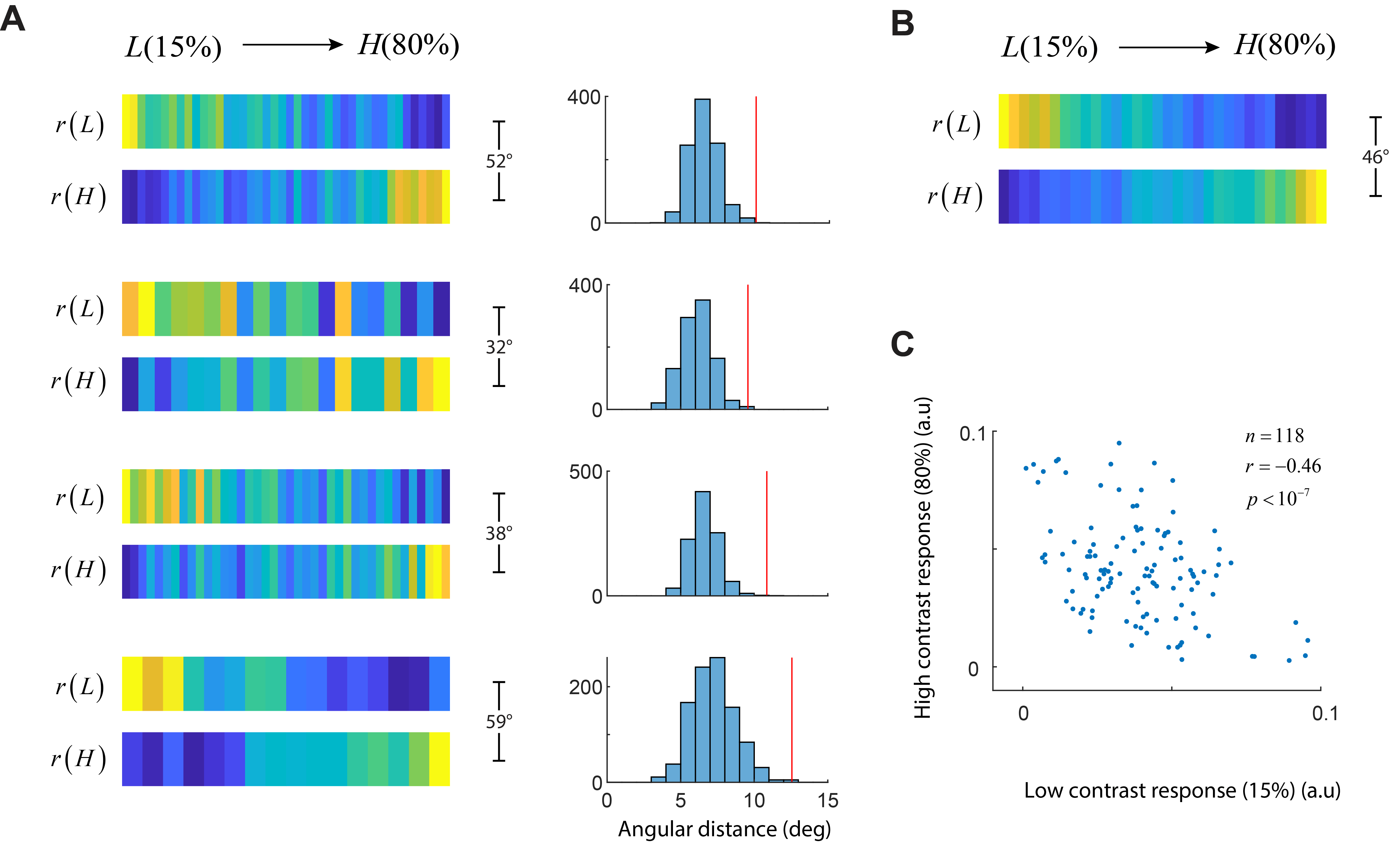}
\justify
\caption{Failure of contrast invariance at the population level. (\textbf{A}) Barcodes to the response of a grating with low (15\%) and high (80\%) contrast in four separate experiments.  In all cases, the angular distance between the population vectors is higher of what is expected by chance. The histograms show the distribution of angular distances in random permutation tests and the red vertical bar shows the angular distance corresponding to a significance level of 0.01. In all cases the angular distance between the high and low contrast is statistically significant.  (\textbf{B}) Average population responses show that some cells (those located on the left of the barcode) are suppressed by increasing stimulus contrast.  (\textbf{C}) There is a negative correlation between responses to high and low contrast.  Cells with a large response at 15\% tend to be suppressed by a stimulus with 80\% contrast, while cells that have a low response at 15\% tend to have a higher response at 80\% contrast. 
}
\end{figure}

\section{Discussion}

Understanding how populations of neurons respond to natural stimulation is a central problem in systems neuroscience \cite{RN1348,RN2214}. Here we studied subspace invariance of population response, which is a weaker property than the additivity and homogeneity of linear systems, as it does not require strict superposition of the responses. In a linear system $r(\alpha A + \beta B) = \alpha r(A) + \beta r(B)$, while while subspace invariance only requires that $r(\alpha A + \beta B) \in \textrm{span}\{r(A),r(B)\}$. One version of the normalization model \cite{RN820} satisfies subspace invariance and appears capable of explaining the responses neural populations in cat primary visual cortex to plaids composed of gratings of different contrast levels. The model accounts for the averaging behavior observed for gratings of equal contrasts and the transition into a winner-take-all regime as the contrasts increasingly diverge, making the identity of the stimulus with highest contrast determine the population response. A different study used intrinsic imaging in tree shrew primary visual cortex and reported averaging behavior when the components had equal contrast \cite{RN2400}. Of course, averaging satisfies subspace invariance as well.  A third model of coding of multiple stimulus involves the temporal multiplexing of their individual responses \cite{RN2436}. According to such model, the response to the joint presentation of two stimuli, $A$ and $B$, will fluctuate over time between $r(A)$ and $r(B)$. This model also satisfies subspace invariance, as the {\em mean} response to the superposition will be linear combination $r(A)$ and $r(B)$. The mixing weights will represent the relative amount of time the population spent in $r(A)$ and $r(B)$. Ruling out subspace invariance, therefore, can potentially rule out several proposals regarding of how neural populations represent sensory information. 

In contrast to past studies in cats and tree shrews, we find gross violations of subspace invariance in mouse primary visual cortex. The responses of neural populations to plaids could not be approximated by the linear combination of the population responses to its components, as evidenced by a large, median angular deviation of 56.5 deg and a relative error of 0.83 (\textbf{Fig 3D \& 4A}). This result falsifies the hypothesis of subspace invariance and, as a result, all models that satisfy such property. The central reason subspace invariance fails is a strong, negative correlation between the average responses cells to the component gratings and to the plaid (\textbf{Fig 5B}). This means there were some cells that responded to the plaids that were unresponsive to the gratings and vice versa.  

Is it possible that subspace invariance holds for sub-threshold responses but fails when analyzing extracellular responses? Let us assume that $r(A)=f(u(A))$, where $u(A)$ is a sub-threshold response to grating A and let $f(\cdot)$ be a monotonically increasing, convex nonlinearity. We also adopt a similar notation for the other stimulus classes. Then, the response to a convex mixture of $A$ and $B$ satisfies $r(tA+(1-t)B) \leq t r(A)+(1-t) r(B)$ for $0\leq t \leq 1$ (Jensen’s inequality). For plaids constructed by averaging of two grating stimuli, we have $t=1/2$ and $r((A+B)/2) \leq (r(A)+r(B))/2$. However, it is clear there are many cells for which this inequality is violated (\textbf{Fig 5B}). Cells with near zero responses to the gratings, should have their responses to the plaid bounded from above by very small number. Instead, these are cells that yield maximal responses to the plaid. Thus, a simple output nonlinearity is an unlikely explanation for the observed phenomenon. (Although our plaid stimuli are not strictly convex mixtures of the component gratings, as the norm of the stimulus constant, it is doubtful our findings would be any different had we chosen to average them instead.)  

In a classic version of the normalization model the response of the $i-th$ neuron in the population is given by $r_i=\langle w_i,x \rangle / \sqrt { (\|x\|^2+\sigma^2) }$ (8, 11). Here, $\langle w_i,x \rangle$ represents the dot product between the receptive field of a neuron $w_i$ and the stimulus $x$, and $\sigma$ is the semi-saturation constant. Assume our stimuli are high contrast, so that 
$\sqrt {\|x\|^2+\sigma^2} \approx \|x\|$. Then the response to $A$ and $B$ are $r_i^A= \langle w_i,A \rangle /\|A\|$ and $r_i^B= \langle w_i,B \rangle / \|B\|$.  
The response to the plaid would be:

\begin{equation*}
r_i^{A+B} = \frac{\langle w_i,A+B \rangle}{\| A+B \|} 
= \frac{\langle w_i,A \rangle}{\| A+B \|} + \frac{\langle w_i,B \rangle}{\| A+B \|} 
= \frac{\|A\| \langle w_i,A \rangle}{\|A\|\| A+B \|} + \frac{\|B\| \langle w_i,B \rangle}{\|B\| \| A+B \|} 
= \frac{\|A\|}{ \| A+B \|} r^A_i + \frac{\|B\|}{ \| A+B \|} r^B_i.
\end{equation*}

As a population, $r^{A+B}=({\|A\|} / { \| A+B \|}) r^A +  ({\|B\|} / { \| A+B \|}) r^B$. In other words, the population response to the plaid ought to be a linear combination of the responses to the individual components.  If the effective contrast (the norm) of the stimuli are the same, $\|A\| = \|B\| = \|A+B\|$, as is the case in the present experiments, one would simply expect the responses to add: $r_i^{A+B}=r_i^A+r_i^B$. Our findings reject such a model and, in general, any model of population coding that satisfies subspace invariance.

Some additional insight into the phenomenon was obtained by analyzing the average dependence of tuning in the joint orientation and spatial frequency domain for cells responsive to either gratings or plaids (\textbf{Fig 6}). Cells that are responsive to gratings but not the plaids show sharp tuning in the orientation and spatial frequency domain. In contrast, cells that respond to the plaids stronger than to the gratings have a broad tuning for orientation and a preference for low spatial frequencies. In our experiments, the power (standard deviation) of the stimuli are kept constant. Thus, while the stimulus power is highly concentrated at one orientation for the gratings, it is distributed across two orientations for the plaid. A more uniform distribution of power across orientations may drive broadly tuned cells to respond better, and as the power at any one orientation is reduced, cells tuned to a specific orientation may experience a reduction of responses to the plaid. This interpretation is consistent with a prior study, showing that single orientations are best represented by populations of highly selective neurons, while orientation mixtures, of which a plaid is a particular case, are best encoded by less selective neurons \cite{RN2437}. The diversity of orientation bandwidths in the cortical population may be beneficial for the coding of natural scene patches, which can exhibit narrow and broad distribution of power across orientations \cite{RN2437}. Thus, subspace invariance fails because stimuli with different orientation distributions engage populations of neurons with different tuning preferences.  

The results uncovered two reasons for the failure of contrast invariance.  First, the responses of some cells in mouse V1 are suppressed by increasing the contrast of a stimulus. Such cells have been observed in the LGN and termed “uniformity detectors” \cite{RN2151}, as they respond best when there is no contrast within their receptive fields (a uniform field). Other cells show the more common pattern of a monotonically increasing response with contrast. A population of cells that includes members from both of these classes cannot be contrast invariant, as the direction of the population vector is bound to change with contrast. The second reason contrast invariance fails at the population level is that, even if we restrict the discussion to cells with monotonically increasing responses, the shape of the contrast response functions will be different across cells. However, as discussed earlier, contrast invariance at the {\em population level} requires all cells to share the same contrast response function. 

There are some technical differences in the stimuli and methods used between this and past studies that may explain some of the differences in our results. We only consider cells which generated a significant response to at least one class of stimuli. Thus, our barcodes, do not contain cells that may be well tuned to intermediate orientations to the ones used in our the stimuli. In contrast, the study by Busse and collaborators measured from cells which responded to gratings at intermediate orientations \cite{RN2215}. Their population responses were defined first by arranging cells according to their preferred orientation and then averaging their responses in a discrete number of orientation bins. This apparently innocuous averaging is important.  First, it will average the contrast sensitivity of a number of cells, making the population appear more contrast invariant that would otherwise be based on the individual responses. A separate study used the actual population responses \cite{RN793} but assessed the model fits using a "quality index" that cannot be directly compared to our estimates of $\varphi$. Second, the analysis may wash out the activities of cells responsive to plaids by averaging them with many other non-responsive cells. Another methodological difference is that recordings by Buse et al were done with a 10x10 micro-electrode array with a grid spacing of $400 \mu \textrm{m}$. Such a population contains cells with pairwise distances much larger than the one in our two-photon data. It is possible that interactions between the members of the population that give rise to the negative correlations observed here are due to local network mechanisms not detectable when sampling from cell pairs which are far apart in the cortex. 

The different results could also reflect true species differences. Although in our experiments stimuli were flashed gratings and plaids, we note that when probed with moving stimuli the proportion of component versus pattern cells in mouse appears different than those cats and monkeys \cite{RN2396,RN2398,RN729} (but see \cite{RN1856} for a different outcome). According to one study \cite{RN2396}, the fraction of component cells in mouse V1 is only 17\%, compared to 84\% in higher mammals, while the fraction of pattern cells is about 10\%, compared to <1.3\% in cats and monkeys. It would be of interest to see if there is any relationship between the responses of neurons to flashed gratings/plaids and the classic definition of component and pattern cells defined by moving stimuli \cite{RN729,RN761}.

Altogether, our data indicate the population responses in mouse V1 to the simultaneous presentation of two stimuli are not readily explained by a linear combination of their responses to the individual components. Moreover, the population vector also changes with contrast, demonstrating a failure of contrast invariance. In other words, spatial structure and contrast interact in a complex way at the population level. More detailed models of population coding which include a realistic diversity of tuning of cortical cells need to be developed to explain the responses of cortical populations.


\section*{Acknowledgements}
We thank Luis Jimenez for help with data collection. This work was funded by NIH grants NIH EY018322 and EB022915 to Dario L. Ringach.

\bibliography{brainbib}

\vfill
\section*{Appendix}

Original review appears in \verb|\normalfont| and authors' replies in \verb|\itshape|.

\hfill

\itshape
We thank the reviewers for their constructive comments.

To summarize, the following are the major changes to the revised version of the manuscript:

1. Following the reviewers’ suggestion, we conducted additional experiments and analyses to study the differences in the tuning properties of cells that respond to gratings vs plaids (new Fig 6).  

2. Following up a question from Reviewer 1, we conducted additional experiments and analyses to test contrast invariance (new Fig 7).

3. Following Reviewer 1 recommendation, we conducted random permutation tests to assess the expected size of angular distances (histograms in Fig 3A and 7).

4. Expanded the introduction to better explain the relationship between space and contrast invariance.

In what follows, we provide a point-by-point reply to each individual comment.
We thank the reviewers again for their feedback.  We hope the revised manuscript has been strengthened as a result of these changes.  
\normalfont

\paragraph{Reviewer \#1}

In this well-written paper, Tring \& Ringach document a previously unknown property of mouse primary visual cortex: subspace invariance does not hold. The experimental measurements are straightforward and yield compelling evidence for this claim. The authors propose that an early nonlinearity may be the cause of this behavior. I enjoyed reading the paper a lot, but feel that there is room for further improvement. First, I think the authors undersell the importance of their findings. Second, I think the statistical analysis needs a bit of work to be more interpretable. Third, I believe their proposed mechanism predicts that tuning in mouse V1 is not contrast-invariant, which would be of major importance if true.

\itshape
Thanks for considering the work potentially important.  Our replies to the recommendations follow the elaboration of these points below.

\vspace{6pt}
\normalfont 

(1)	Importance of these findings I think the introduction should be a bit bigger in scope. Variants of the experimental paradigm used by the authors have been used many times to study mechanisms of gain control in cat and monkey V1. There is a wealth of evidence that divisive normalization explains V1 responses well in these model systems, both at the single neuron level and at the level of populations. Divisive normalization also accounts for many perceptual phenomena in human observers. As the authors explain, normalization is one example of a mechanism that obeys sub-space invariance. If it turns out that the visual cortex of mouse fails to exhibit this behavior, it puts into question how useful this model system is to study basic visual mechanisms, especially as far as understanding the neural basis of human vision is concerned. I would like the authors to expand the scope of the introduction and share their thoughts on this issue.

\itshape
We have modified the introduction and discussion to be more specific on what the results mean. In particular, we state the data rule out a particular form of the normalization model.  We now make this clear in the discussion.

\vspace{6pt}
\normalfont

(2)	Statistical analysis Figure 3B–D is very hard to interpret without any kind of null-hypothesis. Angular distance is a statistic that will always yield a positive value. This makes it hard to judge whether the measured values are truly different from what we would expect if there were no difference between the mean population vectors other than measurement noise. To construct a null-hypothesis, the authors could for example create synthetic data-sets by bootstrapping their data (with replacement), and then compute the angular distance between mean population vectors elicited by the same stimulus condition. Because of measurement noise, these distances will be larger than zero. If repeated many times, this would yield the right reference distribution to compare the cross-stimulus distances with. Figure 4A suffers from a similar problem.

\itshape
Following this suggestion, we have performed random permutation tests to compute the distribution of angular distances under the null hypothesis that there is no relation between the population responses and stimulus classes.  These distributions, along with the values of angular distances at the 0.01 significance level, are now shown in Fig 3 and 7.

\vspace{6pt}
\normalfont

(3)	Proposed mechanism I think the proposed mechanism is interesting and plausible. I would like the authors to develop it a little further. Please simulate the responses of orientation selective mechanisms and show that the resulting population responses approximate your observations when proceeded by a saturating non-linearity. Isn’t the implication of this mechanism that you predict that the outcome of this experiment would be very different at lower contrasts? If the stimulus contrast is too low to elicit response saturation, the distortion should not happen, and subspace invariance should hold. In other words, mouse V1 would not exhibit contrast-invariant tuning for plaids. It seems worthwhile to spell out this prediction, and test it in future experiments. Contrast-invariance is another core property of cat and monkey V1, and is thought to be essential for interpreting the content of the visual environment by areas downstream of V1.

\itshape
We have dropped this section as it became clear from the new data on the tuning of cells, collected in response to a request by Reviewer \#2, that something different is going on (Fig 6). Namely, plaids seem to recruit the activation of cells that are more broadly tuned in orientation and prefer lower spatial frequencies. 

In addition, to address the point raised about contrast invariance we now expanded the introduction to explain the relationship between subspace invariance to contrast invariance (as pointed out by the reviewer it is a special case) and we also collected some additional data that shows the failure of contrast invariance (new Fig 7).

\vspace{6pt}
\normalfont

Minor things
(1) The methods mention a couple of things that need either more or less. Locomotion and eye movements were monitored. What was done with these measurements? How do they relate to the data analyzed in this paper?

\itshape
These were ancillary measurements in our experiments but are not used in the present analyses.  We have removed this description from the text and methods Fig 2.

\vspace{6pt}
\normalfont

(2) Spikes were estimated via deconvolution. Is anything done with inferred spikes in this paper? That was not clear from the rest of the text. 

\itshape
Yes, all responses and their analyses are based on the estimated spiking after deconvolution. We have now clarified this in the text.
\normalfont

(3) It is not clear what the different curves in Fig. 2C represent 

\itshape
The different curves represented the responses of different cells.  To remove the clutter we now show just the response of a single neuron.

\vspace{6pt}
\normalfont

(4)	How are the barcodes ordered exactly? 

\itshape
The ordering of cells in the barcodes is described in lines 151-157.  They are ordered according to the difference in responses to the two grating stimuli.

\vspace{6pt}
\normalfont

(5)	The ordering never seems perfect, but it is not clear from the text why that is the case. 

\itshape
We are not sure what “perfect” means.  Note that the ordering is with respect to the difference in firing to A and B, but that the barcodes show the responses to A, B and C=A+B.  Thus, cells in the middle of the barcode respond equally well to A and B but their firing rates can be equally low or equally high. We now clarify this point in the text (line 153-157).

\vspace{6pt}
\normalfont

(6)	Why are the angular distances between r(A) and r(B) not closer to 90 deg?

\itshape
Because there is a group of cells respond to both A and B.  So even if stimuli are orthogonal they do not necessarily evoke orthogonal responses 

\vspace{6pt}
\normalfont

\paragraph{Reviewer \#2}

In this study, the authors measure population responses in mouse primary visual cortex to single gratings and plaids, and report that the population response to plaids cannot be explained by a linear combination of the responses to the individual components. While the question of subspace invariance is very interesting and relevant, I am concerned that the results presented in this paper seem to fall short on depth: as discussed by the authors, interesting insights could be obtained by knowing the orientation preference of the recorded neurons. 

\itshape
Thank you for this important suggestion. We followed the reviewer’s recommendation and collected additional data to measure the tuning in both orientation and spatial frequency of cells that respond preferentially to the grating and plaid stimuli (new Fig 6).

\vspace{6pt}
\normalfont

Besides the general issue of depth, I am worried about a technical point: the authors should rule out that the ‘early nonlinearity’ might be generated in the stimulus presentation instead of the early visual system.

\itshape
It is a reasonable concern.  As stated in the methods section, we carefully linearize our monitors using a PR-655 spectro-radiometer (lines 91-94).  We also verify the linearity of the monitor after the look-up tables are computed and installed.  Thus, we are confident there are no non-linearities generated at the monitor. Moreover, as explained above, the proposed interpretation of the data as resulting from an early non-linearity has been dropped due to the insights provided by the measurement of the tuning kernels as suggested by this reviewer (new Fig 6), which suggests a different explanation for the phenomenon.

\vspace{6pt}
\normalfont

MAJOR: Mouse vision has low acuity. Can you please provide a realistic illustration of what the sinusoidal plaid looks like on the monitor, and where approximately the RFs of the recorded neurons are. 

\itshape
As noted in the text “The center of the monitor was positioned with the center of the receptive field population for the eye contralateral to the cortical hemisphere under consideration. The approximate locations of the receptive fields of the population were estimated by an automated process where localized, flickering checkerboards patches, appeared at randomized locations within the screen. This experiment was run at the beginning of each imaging session to ensure the centering of receptive fields on the monitor. We imaged the monocular region of V1 in the left hemisphere. The receptive fields of neurons were centered around 20 to 35 deg in azimuth and 0 to 20 deg in elevation on the right visual hemifield.” (lines 106-113).
\normalfont

The stimuli used for illustration seem to have too high spatial frequencies. How many cycles of the stimulus will fall into the RFs? Given that the spatial frequencies are so low, can you please exclude that the early non-linearity is in the visual stimulus itself? For example, what was the bit-depth used to generate the visual stimuli? Which additional non-linearities are introduced by the LED screen? Could the angle with which the monitor was viewed add additional distortions?

\itshape
The typical spatial frequency of the stimuli was 0.04cpd, which for a screen 112deg wide, translates to ~4.5 cycles per screen. Fig 1 was modified to show this typical scenario.  Because the screen is placed normal to the line of sight, the receptive fields centered on the screen, and the entire population has RFs within the central +/- 20 deg, there are no major changes in luminance due to the viewing angle of the monitor or major perspective distortions.

\vspace{6pt}
\normalfont

To make the subspace invariance part more accessible to the general audience, can you please extend the explanation at the end of the introduction? Related to Fig. 1, could you please state exactly what “population response” means in this context? In Fig. 1, should the axes represent this population vector pointing in some direction and with some length? If yes, how is the population vector computed? Later, it’s said that the dimension of the trial-by-trial population response to a single stimulus equals the number of cells in the population. How does this fit to the illustration in Fig. 1?

\itshape
A population response is a vector where the i-th entry contains the mean response of the i-th cell to a stimulus (see also description in lines 145-148 and caption to Fig 1). The dimension of the vector is the number of cells in the population. Fig 1 is a schematic showing the mean responses to each stimuli class as a vector. Note that population responses represent the mean across trials.

\vspace{6pt}
\normalfont

How does the computed angle between the population vectors depend on the number of neurons and their orientation preferences? I understand that orientation tuning was not (yet) measured, but some of these questions might be answered by simulations?

\itshape
We now measured the orientation and spatial frequency tuning of cells and how these relate to their responses to gratings vs plaids (new Fig 6).  We have no studied in detail how the number of neurons in a population affect the expected angles.  Of course, one would expect angular distances to increase with the \# of cells.  Note, however, that the new calculations from random permutation tests take the dimension of each population into account.
\normalfont

Can you please add information on how the “barcode” was normalised? From the examples shown in Fig. 2, it might look like the responses to the individual conditions were normalised separately. Do the results also hold if the normalization is performed across conditions?

\itshape
Each barcode is normalized between zero and its maximum response separately. We now state this in the legend of Fig 3. This is merely done for visualization purposes.  Note that angular distances are independent of how these are normalized.  Moreover, had we normalized the barcodes altogether, the Figures would have changed little. Indeed, Fig 4B and 5B show that the magnitude of the response to plaids and gratings were similar. 

\vspace{6pt}
\normalfont

p. 4: can you please explain the rationale of providing the same “effective contrast” for the plaid? Most classic studies have related the response of individual gratings to just their sum without normalising by square root of 2. This is also related to one sentence in paranthesis in the discussion session.

\itshape
Contrast gain control in the early visual system is determined by the standard deviation of the contrast of the stimulus (see http://www.jneurosci.org/content/26/23/6346.short) By keeping the norm of the different stimuli constant, we ensure that the gain control of cells in the population remains at a constant level during the recordings. The normalization by sqrt(2) keeps the standard deviation of the plaid equal to that of the gratings. 

\vspace{6pt}
\normalfont

The discussion reads very fragmented. E.g., on page 9, last paragraph, there is a piece on comparison to Busse et al. (2009) that seems to be taken out of context.

\itshape
Fixed.

\vspace{6pt}
\normalfont

Overall, the paper would gain in depth if it included the measurements of orientation preference of the neurons, as discussed by the authors.

\itshape
We now provide such measurement (new Fig 6).

\vspace{6pt}
\normalfont

MINOR:

•	Please add line numbers to help the reviewer

\itshape 
Done.

\vspace{6pt}
\normalfont

•	Fig. 1: Is this illustration meant for the single neuron case?

\itshape 
No, Fig 1 represent population vectors.

\vspace{6pt}
\normalfont

•	p. 3: please describe in more detail the cranial window that was used for imaging - the only aspect mentioned is an “aluminum bracket”

\itshape 
This is really a standard preparation in the field. We have now cited one of the original papers for reference.
\normalfont

•	p. 4: “peak relative change in fluorescence over was larger” - word is missing

\itshape 
Thank you. Fixed.

\vspace{6pt}
\normalfont

•	Fig. 2: It seems like some neurons have both on and off responses. Do the results hold if you only consider the on response?

\itshape
The present analyses include all cells, but yes – results hold if one considers only ON responses.

\vspace{6pt}
\normalfont

•	p. 6: how was statistical differences in the distributions of population angles determined?

\itshape 
Rank-sum test (we have added this information to the text).

\vspace{6pt}
\normalfont

•	p. 6: reference to Fig. 2B should be 4B

\itshape
Thank you!  Fixed.
\normalfont

\end{document}